\documentclass[sigconf]{acmart}

\AtBeginDocument{%
  \providecommand\BibTeX{{%
    \normalfont B\kern-0.5em{\scshape i\kern-0.25em b}\kern-0.8em\TeX}}}

\setcopyright{rightsretained} 
\acmConference[EARS'19]{SIGIR 2019 Workshop on ExplainAble Recommendation and Search}{July 25, 2019}{Paris, France}

\begin{document}

\title{Assessing the Helpfulness of Review Content for Explaining Recommendations}


\author{Diana C. Hernandez-Bocanegra}
\affiliation{%
  \institution{University of Duisburg-Essen}
  \city{Duisburg}
  \country{Germany}
}
\email{diana.hernandez-bocanegra@uni-due.de}

\author{Jürgen Ziegler}
\affiliation{%
  \institution{University of Duisburg-Essen}
  \city{Duisburg}
  \country{Germany}
}
\email{juergen.ziegler@uni-due.de}

\renewcommand{\shortauthors}{Hernandez-Bocanegra and Ziegler}

\begin{abstract}
  Despite the maturity already achieved by recommender systems algorithms, little is known about how to obtain and provide users with a proper rationale for a recommendation. Transparency and effectiveness of recommender systems may be increased when explanations are provided. In particular, identifying of helpful argumentative content from reviews can be leveraged to generate textual explanations. In this paper, we investigate the reasons why a review might be considered helpful, and show that the perception of credibility and convincingness mediates the relationship between helpfulness and the perception of objectivity and relevant aspects addressed. Our findings led us to suggest an argument-based approach to automatically extracting helpful content from hotel reviews, a domain that differs from those that best fit classical argumentation theories.
\end{abstract}

\ccsdesc[500]{Information systems~Recommender systems}
\ccsdesc[500]{Human-centered computing~User studies}

\keywords{Recommender systems, user study, explanations, argument mining}


\maketitle

\section{Introduction and related work}
Recommender Systems (RS) have become widespread tools that aim at facilitating users’ search and decision making in a large variety of domains such as e-commerce or media streaming. . While the algorithmic accuracy of RS has been improved considerably over the years, from a user perspective, most systems act as black boxes that provide no rationale for their decisions and that do not allow users to control or question the recommendations given \cite{Herlocker00}. Research in RS \cite{Tintarev12} has shown that providing explanations can be an effective means for increasing transparency, facilitating users’ decision-making  and increasing user satisfaction as well as the trustworthiness of the system . In particular, textual explanations can provide valuable information that helps users to make better decisions \cite{Costa18}. In this regard, the interest in exploiting online reviews in RS has increased in recent years, given the richness of its content on aspects of interest to the user, not only to improve the accuracy of RS predictions \cite{Catherine17,Zheng17,Ling14,McAuley13}, but also to provide textual explanations of the recommended items \cite{Costa18,Zhang2014a}. For example, \cite{Zhang2014a} proposed a matrix factorization based model to leverage sentiment analysis on aspects addressed in reviews, to provide both recommendations and textual explanations based on templates. \cite{Zheng17} proposed a deep learning approach, using two parallel neural networks to simultaneously model items and users from reviews, in order to predict ratings. This approach was extended by \cite{Chen18}, who stressed that not all reviews should contribute in the same way to item modeling, and that predictions could be improved by using reviews that were useful to the user. For this purpose, an attention mechanism was used, demonstrating that this could not only improve the accuracy of predictions compared to approaches that consider all reviews equally, but also select relevant information that could be used to provide explanations. Despite the convenience of such a concept, \cite{Chen18} determines usefulness by only addressing user's aspects of interest. In contrast, there are many additional features that can also contribute to usefulness.

Extensive research on review helpfulness prediction has been highly focused on content features (eg. review length, readability or word-based features) \cite{Liu07, Yang15, Kim06}. However, helpfulness could also be predicted by using features that convey convincing content and provide hints to credibility. In this sense, and given its persuasive nature, an argument-based approach could help to identify this type of content, through the understanding of underlying patterns and the argumentative-type features contained in reviews, bearing in mind the impact that appropriate argumentation has on the beliefs and behaviors of consumers towards an item. Therefore, we believe that the exploiting of arguments given by users in reviews can contribute to helpfulness prediction. Nevertheless, little is known about whether features related to arguments can work as a good predictors of the helpfulness. In this regard, \cite{Liu17} coined the term “argument-based features” (e.g ratio between claims and premises or position of the argument components), examined and annotated argument components in reviews of hotels domain, and found that when argument-based features are used, helpfulness prediction performance is increased over the use of traditional content features. Despite these findings, it is important to note that there is still a lack of consensus on whether the reviews reflect argumentative structures, not to mention the lack of reliable annotated corpus that would facilitate the automatic extraction of such a content \cite{Habernal17}, not being the case for domains such as legal documents and debates, in which computational argument mining has made significant progress in recent years. 

Overall, we describe in this paper our approaches to explainable recommendations, focusing on the extraction of arguments from reviews. We hypothesise that the use of helpful reviews with convincing and credible arguments that refer to aspects of interest to the user can improve not only rating predictions, but also be the base to generate useful textual explanations. As a first step, we seek to assess in this paper what makes a review be perceived as helpful. Accordingly, we present a user study in which users were asked to rate the helpfulness of a series of reviews and to report their reasons for voting. Here, our specific hypothesis is that features related to arguments (i.e. balance between pros and cons, opinions supported by facts, and a stringent flow of arguments) can work as good predictors of helpfulness in the hotel reviews domain, as well as coverage of aspects relevant to users, and even better than other traditional content features, i.e. review length and the level of detail. Lastly, we propose an argument-based approach to extracting helpful content from reviews, based on our findings.

\section{Helpfulness study}
We hypothesized that features related to arguments can work as good predictors of helpfulness in the hotel reviews domain, in addition to the addressing of aspects relevant to users, and better than other traditional content features, i.e. review length and level of detail. 

\subsection{Method}
We conducted a study survey with 108 participants (49 female, mean age 34.38 and range between 19 and 68) recruited through Amazon Mechanical Turk. All participants were requested to read and rate helpfulness for the same set of 18 hotel reviews.

{\itshape Procedure:} We randomly selected 18 reviews for three hotels in the San Francisco area.  For this, we used the ArguAna dataset for argumentation analysis in hotels domain \cite{Wachsmuth14}, which includes annotations of positive and negative opinions, neutral statements, and aspects addressed. We asked the participants to read 6 reviews for each of the 3 selected hotels, presented at random, and to rate them according to how helpful they considered the review was, and the reasons that contributed to that perception. We asked validation questions at the end of each set of hotel reviews, in order to ensure that participants were actively reading the content. No information of the overall sentiment rate given by author or further information about her was given, seeking for participants to only evaluate the text contained in the review.

{\itshape Questionnaire:} The review text was displayed, followed by the question “How helpful was this review?” and a 1–5 Likert-scale for the response (1: Not helpful at all, 5: Very helpful). Next, participants were requested to inform their opinion about 10 different reasons for their helpfulness reply, in relation to: 1) argument-based features, which are the focus of this study: {\itshape objectivity} ("the review includes an adequate amount of objective statements based on facts"), {\itshape pros and cons} ("the review provided a balanced view of pros and cons") and {\itshape flow of arguments} ("the review has a stringent flow of arguments"); 2) content-based features: {\itshape length} ("the review was too short or too long") and {\itshape level of detail } ("the level of detail provided was too little / too much"); 3) {\itshape aspects} ("the review addresses the aspects that are relevant for my purposes"); 4) perception of {\itshape credibility} ("the review seems credible") and {\itshape convincingness} ("the review provided convincing reasons"); 5) content that is {\itshape emotional} ("the review contains emotional content") or {\itshape episodic} ("the review contains information that might be only episodic"). Opinions were rated on a 1–5 Likert-scale (1: Strongly disagree, 5: Strongly agree).

\subsection{Results and Discussion}

Multiple logistic regression analysis was used to test the reasons that significantly predict participants' ratings of helpfulness. Here, positive coefficients would indicate that the reason contributes to an increase of helpfulness perception, whereas negative coefficients would indicate the opposite effect. The results of the regression indicated the 10 predictors explained 25.4\% of the variance and the model predicts 74.5\% of the responses correctly. The Pseudo-R\textsuperscript{2} is 0.23.  As expected, credibility significantly predicted helpfulness ($\beta$= .29, p<.001), as did aspects. Results also revealed that an increase in the variable length leads to a decrease in the perception of helpfulness ($\beta$=-.20, p<.01). In contrast to \cite{Yang15}, this finding confirms our expectation that long reviews are not necessarily considered helpful, and that other reasons related to argumentative content can have a positive impact on the perception of helpfulness, at least in hotels domain. Such is the case with the variables objectivity ($\beta$=.23, p<.01) and convincingness ($\beta$=.23, p<.01). In the other hand, contrary to our expectations and previous findings \cite{Connors11}, pros and cons is not strongly correlated to helpfulness , nor is flow of arguments. The latter might suggest that even if helpful reviews involve some sort of argumentative content, they do not necessarily follow a strict flow of arguments; in other words, it could be inferred that the presentation of a list of arguments is sufficient, and that a stringent flow of arguments is not really relevant to the perception of helpfulness in this domain. On the other hand, the level of detail not only is not a significant predictor of helpfulness, but we found that is highly correlated to length (r\textsubscript{s}=.69, p<0.001), reason why we decided to exclude it from further analyses. Additionally, we found that emotional is also not a significant predictor of helpfulness; even when some of the reviews than can be read online appeal to emotions in order to persuade, this may be a dimension that does not apply transversely to most types of users, as if it could be the case of objectivity. Lastly, we consider that, even when episodic is not a significant predictor of helpfulness, is an interesting aspect that should be addressed further. We consider that content related to an incident that could be perceived as irrelevant in a single review -but appears repeatedly in subsequent reviews- could change the perception of helpfulness, once several reviews are read. Table 1 shows the coefficients obtained for all the model variables.

\begin{table}
\caption{Logistic regression of reasons for voting on helpfulness}
\label{tab:conf}
\begin{minipage}{\columnwidth}
\begin{center}
\begin{tabular}{lccc}
\toprule
Reason&Coefficient\footnote{(*) p<0.05, (**) p<0.01, (***) p<0.001}&p value&Odds ratio\\
\midrule
Length&-0.200**&0.00&0.82\\
Level of detail&-0.108&0.11&0.90\\
Objectivity&0.232**&0.00&1.26\\
Pros and cons&0.007&0.91&1.01\\
Convincing&0.225**&0.00&1.25\\
Aspects&0.228**&0.00&1.26\\
Flow of Arguments&0.024&0.71&1.02\\
Credibility&0.293***&0.00&1.34\\
Emotional&0.048&0.42&1.05\\
Episodic&-0.074&0.22&0.93\\\\
Accuracy: &74.5\%\\
Pseudo-R\textsuperscript{2}: &0.23\\
\bottomrule
\end{tabular}
\end{center}
\end{minipage}
\end{table}

In addition, we wanted to understand in more detail the mechanisms underlying the relationship between helpfulness and observed variables that can be considered ‘less subjective’ (in contrast to variables like credibility and convincingness), since they can be quantified through the use of complementary information. (e.g. participants were asked to inform their aspects of interest, and reviews are already annotated with aspects, facts and opinions). Credibility and convincingness could, on the other hand, be evaluated as mediators of the relationship between helpfulness and some of the rest of observed variables. To this respect, two mediation models were tested in three steps according to standard procedure \cite{Baron86}. First, we regressed helpfulness on all independent variables but credibility and convincingness, and we confirmed that the length ($\beta$=-.18, p<0.01),  objectivity ($\beta$=.28, p<0.001) and aspects ($\beta$=.31, p<0.001) are significant predictors of helpfulness. Second, we regressed the mediator credibility on the independent variables, and confirmed that objectivity ($\beta$=.14, p<0.05) and aspects ($\beta$=.18, p< 0.01) are significant predictors of the credibility. The same was the case with convincing as mediator, and confirmed that details ($\beta$=-.19, p<0.01), as well as objectivity ($\beta$=.17, p<0.05) and aspects ($\beta$=.39, p<0.001) are significant predictors of convincingness. In the third step, the association of aspects and objective and helpfulness reduced significantly when credibility and convincingness were added to the model. Therefore, credibility and convincingness act as mediators of the relationship between helpfulness and objectivity and relevant aspects addressed (Figure 1). The foregoing means that credibility and convincingness serve to clarify the nature of the relationship between helpfulness and aspects and objectivity. Thus, addressing relevant aspects and objectivity influences credibility and convincing, which in turn influence helpfulness.

\begin{figure}[h]
  \centering
  \includegraphics[width=\linewidth]{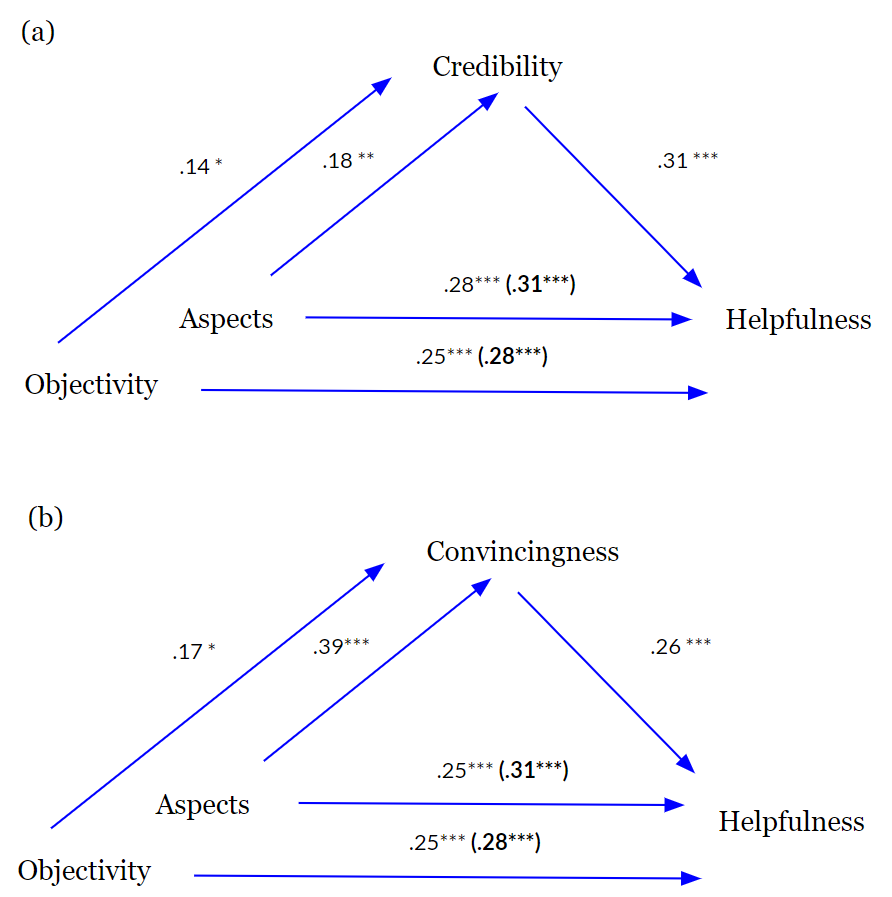}
  \caption{Credibility (a) and convincingness (b) as mediators between helpfulness and aspects and objectivity. Coefficients in bold represent the strength of linear relation between helpfulness and aspects and objectivity, when credibility and convincing are not statistically controlled by including them as predictors of helpfulness.}
  \Description{Mediation analysis}
\end{figure}

\section{AN ARGUMENT-BASED APPROACH TO EXTRACTING HELPFUL CONTENT FROM REVIEWS}

Even when classical argumentative structures are not representative of the content normally found in online reviews, the authors offer, in many cases, valuable information of persuasive purpose. In this sense, and based on the findings of our study, our proposed approach is based on the idea that objective statements supported by facts provide very helpful information to users. In consequence, helpful content could be extracted by means of a shallow argumentative structure, that represents objective reasons or evidences that support an opinion on a certain aspect. A similar approach is used by \cite{SaintDizier12}, where an argument is represented by ‘reasons supporting a conclusion’, although no distinction is made between objective and subjective reasons. On the other hand, we define objective reasons as statements that include facts, and subjective opinions as statements with polarity in respect to an aspect, e.g. “Don't count on the breakfast that is included [opinion, polarity: negative]; instant coffee and packaged muffins, etc [fact]” For the automatic extraction of these components, we plan to use the ArguAna corpus \cite{Wachsmuth14}, which includes a statement-level annotation of facts, positive and negative opinions, and aspects for hotel reviews extracted from TripAdvisor (inter-annotator agreement Fleiss' $\kappa$=0.67).  In spite of the convenience of this corpus, the annotation scheme needs to be complemented in order to useful for our purposes. In particular, the annotation of facts that relate to properties and services provided by the hotel is still needed, since neutral statements from ArguAna include not only facts, but all the statements without polarity, like contextual information (e.g. “We stayed here New Year's Eve”), or suggestions and tips that are not directly related to the evaluated hotel (e.g. “We found a great Japanese restaurant called Wasabi and Ginger on Van Ness Avenue just 3 mins walk away - highly recommended”). In addition, the support relation between opinions and facts should also be annotated, as a basis for the automatic relation detection task. 

In addition, we examined the content of reviews with the best helpfulness scores in our user study (section 2), and found a number of argumentative figures that represent potentially convincing content (Table 2). In the future, a detailed analysis is required to establish their impact on helpfulness, the feasibility of its annotation and automatic extraction, and to what extent its use would improve the quality of the extracted arguments. 

\begin{table}
\caption{Examples of additional argument figures found in reviews perceived as helpful}
\label{tab:conf}
\begin{minipage}{\columnwidth}
\begin{center}
\begin{tabular}{lll}
\toprule
Figure&Aim&Example\\
\midrule
Match&Establish a contrast or a&“you get what you\\
reality vs.&match between initial& pay for.”\\
expectation&expectation and the real&\\
&perception of author.&\\
&&\\
Moderately &Establish nuances of&“A low-cost place\\
negative/&polarity, which may reflect&to sleep with no\\
positive &a honest author’s opinion,&frills.”\\
statements&and consequently increase&\\
&the perception of&\\
&credibility.&\\
&&\\
Episodic by&Enumerate examples to&“The staff was very\\
example&support an opinion. In this&friendly and\\
&case, unlike facts like&helpful. For\\
&breakfast is included,&example, they set\\
&we refer to events that&up transportation\\
&could be episodic.&for me on multiple\\
&&days without a\\
&&hitch.”\\
&&\\
Preferential&Ilustrate opinions that&“Room was clean\\
&reflect author’s personal&and the place was\\
&preferences, that involve a&nice. But I enjoyed\\
&comparison with similar&some of the other\\
&items.&Joie De VIvre\\
&&hotels in the same\\
&&two block area\\
&&more.”\\
\bottomrule
\end{tabular}
\end{center}
\end{minipage}
\end{table}

We present above our approach to extracting helpful arguments from reviews. However, our aim is, in a broader sense, to use such content not only to improve rating predictions, but also to generate helpful textual explanations that reflect system transparency and facilitate users decision making. Therefore, in future work, we plan to develop and integrate methods that:

\begin{itemize}
\item Facilitate the automatic extraction of helpful arguments from reviews. We plan to base our work on developments from argument mining \cite{Eger17,Habernal17,Schultz18} (for the supervised detection of units and its relationships on a multi-sentence level), sentiment and subjectivity analysis (for detection of objective/subjective units on a sentence level \cite{Radford17}).
\item  Establish the relevance of the extracted arguments to generate proper explanations and improve rating prediction, by leveraging attention mechanisms \cite{Bahdanau14} to detect aspects of interest to user.
\item  Allow to generate natural language explanations of recommended items, using the relevant arguments extracted. Here, we plan to base our work on abstractive summarization techniques as proposed by \cite{Nallapati16}.
\end{itemize}

According to our initial approach, all reviews containing arguments (subjective opinions supported by objective facts) that are relevant to users’ interests would be considered helpful and used as the basis for RS predictions and explanations. This binary approach (review is helpful or not) could be extended to determine different degrees of helpfulness, so that higher quality arguments have a greater weight in the rating prediction. In this respect, the definition of additional criteria to improve the evaluation of the relative quality of arguments is still needed; however, the use of features such as the ratio between facts and opinions, or the length of sentences that address aspects relevant to users could be a starting point for this purpose.

Furthermore, we also plan to address in the future the applicability of the proposed approach to domains other than hotel reviews, e.g. restaurant or product reviews.
\section{CONCLUSIONS}
In this paper we have discussed the use of online reviews, in particular to generate explanations that can serve the objectives of transparency and effectiveness of RS. A novel way of extracting helpful content from reviews was proposed, based on a arguments-based approach. To support our proposal, we presented a user study, which sought to establish the reasons why a user might find a review helpful. Our findings lead us to suggest that arguments in the form of opinions supported by objective statements provide very helpful information to users. As future work, we aim to implement this concept and use it, not only to improve RS predictions, but also to generate textual explanations of the recommended items.

\begin{acks}
This work is supported by the German Research Foundation (DFG) under grant No. GRK 2167, Research Training Group “User-Centred Social Media”.
\end{acks}

\bibliographystyle{ACM-Reference-Format}
\bibliography{helpfulness-base}


\end{document}